\documentclass[twoside]{article}
\usepackage{amsmath,amsfonts,amssymb,enumerate}
\usepackage{qic,epsfig}
\usepackage{cite}

\newcommand{\ket}[1]{| #1 \rangle}
\newcommand{\bra}[1]{\langle #1 |}

\begin{document}
\setlength{\textheight}{8.0truein}    

 \runninghead{A Criterion for the Bipartite Separability of Bell Diagonal States}
            {}

\normalsize\textlineskip \thispagestyle{empty}
\setcounter{page}{1}

\vspace*{0.88truein}

\alphfootnote

\fpage{1}

\centerline{\bf A Criterion for the Bipartite Separability of Bell
Diagonal States} \vspace*{0.035truein} \centerline{\footnotesize
 Ming-Chung Tsai, Po-Chung Chen,  Wei-Chi Su,  and Zheng-Yao Su\footnote{Email: zsu@nchc.gov.tw}\hspace{.15cm}}
\centerline{\footnotesize\it Department of Physics, National Tsing
Hua University, Hsinchu, Taiwan, R.O.C.}
\centerline{\footnotesize\it National Center for High-Performance
Computing, Hsinchu, Taiwan, R.O.C.} \centerline{\footnotesize\it
National Center for Theoretical Sciences, Hsinchu and Tainan,
Taiwan, R.O.C.}

\vspace*{0.21truein}

 \abstracts{
 A decomposition form is introduced in this report to establish
 a criterion for the bi-partite separability of Bell diagonal states.
 A such criterion takes a quadratic form of the coefficients of a given Bell diagonal states
 and can be derived via a simple algorithmic calculation of its invariants.
 In addition, the criterion can be extended to a quantum system of
 higher dimension.
 }{}{}

 \vspace*{10pt} 
 \vspace*{3pt} \vspace*{1pt}\textlineskip

 \section{Introduction}\label{secintro}
  Entanglement is a type of characteristic that cannot be described
  by the classic physics.
  It is also known as an essential resource to a quantum
  processor and a quantum computation~\cite{}.
  A problem of great importance in the field of quantum information science
  is to qualitatively analyze entanglement.
  That is, determine whether a state
  is {\em separable} or {\em entangled}.
  According to the article of Werner in 1989~\cite{Werner},
  a state of a bipartite system ${\cal H}^{AB}$ is separable if it can be written as a {\em convex} combination of pure product states
 \begin{align}\label{eqdefnsep}
  \rho^{AB}=\sum^N_{i=1}p_i\rho^A_i\otimes\rho^B_i,
 \end{align}
 here $p_i\geq 0$, $\sum^N_{i=1}p_i=1$ and $\rho^{\mathcal{I}}_i$
 being a pure state in a Hilbert space
 $\mathcal{H}^{\mathcal{I}}$ with $(\rho^{\mathcal{I}}_i)^2=\rho^{\mathcal{I}}_i$ for $\mathcal{I}=A,B$.
 Otherwise, this state is {\em entangled}.

 An enormous number of research works have been realized to search for the criteria
 to answer the separability problem of a state.
 The earliest criterion, reported by Peres~\cite{Peres},
 is to use the {\em partial transpose} of a given density operator.
 A such criterion provides a sufficient and necessary conditions of
 deciding the separability of states in lower-dimensional bipartite systems~\cite{Horo2},
 including  $2\times 2$ and $2\times 3$ systems.
 There are other operational criteria for separability,
 such as concurrence criterionn~\cite{Wootters}, reduction criterion~\cite{Horo3}, majorization criterion~\cite{Nielsen}.
 Nevertheless, it is very difficult to examine whether some given states of any dimension can be written as a mixture of product states~\cite{Horo1}.
 On the other hand, both entanglement witnesses and positive maps are sufficient and necessary conditions under any dimension system, but these two criteria are non-operational~\cite{Horo2}. Based on the above-mentioned criteria, some researchers developed various methods to search for optimal measurements on the given density operators~\cite{Doherty,Eisert}.

 In $ 2\times2 $ system, the Bell diagonal states can be characterized by three-parameter set, whose separability are complete studied~\cite{Horo1,Horo4}. The Bell diagonal states in $ 3\times3 $ system are introduced in~\cite{Bau1}, then Baumgartner, et al~\cite{Bau2} extend their study to $ d\times d $ system. In addition to these, more properties of Bell diagonal states were analyzed ~\cite{Bau3}. Numerous attempts have been made by scholars to write down the decomposition form for Bell diagonal states, but most of them are limited to special cases (Werner state)~\cite{Azuma,Unanyan,Milburn,Rungta,Pittenger}. Sanpera, et al.~\cite{Sanpera} utilize a constructive algorithm to decompose the separable state in either a $2\times 2$ or $2\times 3$ system.
 The decomposing procedure is examined by Werner state with a non-unique decomposition.
 Another decomposition method, which are developed by Wootters~\cite{Wootters},
 based on the minimum average entanglement of an ensemble of the eigenstates of a density matrix for Bell diagonal states in a $2\times 2$ system. Although Wootters's method is a successful measure, it is difficult that the physical phenomena to observe when the separability of Bell diagonal states are transformed into entanglement.

 In this article, we focus on the separability properties of Bell diagonal states
 in a $2^p\times 2^p$ system, We propose a criteria (necessary condition) for the bi-partite
 separability of the Bell diagonal states in a $2^p\times 2^p$ system,
 and write down a new {\em separable form} for the Bell diagonal states in a $2\times 2$ system,
 which is different from the convex combination obtained by previous research work.
 In order that the any $d\times d$ systems $(2^{p-1}<d\leq2^{p})$ could be analysis,
 they can be embedded to the $2^p\times 2^p$ system.
 This research work is organized in the following ways.
 In sec.II we review the relation between the standard basis and the spinor basis (identity matrix and Pauli matrices).
 In the spinor basis, we could obtain a necessary condition
 of the separable Bell diagonal states base on the inequality $Tr^2(\rho)\geq Tr(\rho^2)$.
 Besides, we carry out the proof of the sufficient condition via
 presented decomposition for the bi-partite
 separability of the Bell diagonal states in a $2\times 2$ system.
 This process, based on the definition of density operators (unit trace, hermitian, and positive-semidefinite),
 is not only different from the method used in~\cite{Azuma,Horo2,Wootters,Milburn},
 but gives us an insight into quantum entanglement.
 When the separability of Bell diagonal states are transformed into entanglement,
 implied the local density operators $\rho^{\cal I}_k$
 moved to the outside the Hilbert space ${\cal H}_{\cal I}$ for ${\cal I}=A,B$.
 Then, we would operated Peres PT criterion and compared this result with presented.
 It is known on the condition $p=1$ that $\rho_{B}$ is separable iff $1\leq\sum^3_{i=1}|\Omega_{ii}|$.
 In sec.III we extend this schemes to the condition $p>1$ and acquire the inequality

 \section{Bell Mixture in a $2\times 2$ System}\label{secBell2by2}
  The discussion begins with maximally entangled states in the simplest bipartite system,
  a two-qubit system.
  The formulation of a density operator in this article
  employs the {\em spinor representation} such that one can {\em recursively}
  extend the scheme designed in a two-qubit system to that in a multi-qubit system.

  By definition, the four maximally entangled two-qubit states(Bell's states) could be expressed as:
  \begin{align}\label{eq-Bellsstates}
  \ket{\Phi^+}=\frac{1}{\sqrt{2}}(\ket{00}+\ket{11}),\notag\\
  \ket{\Phi^-}=\frac{1}{\sqrt{2}}(\ket{00}-\ket{11}),\notag\\
  \ket{\Psi^+}=\frac{1}{\sqrt{2}}(\ket{01}+\ket{10}),\notag\\
  \ket{\Psi^-}=\frac{1}{\sqrt{2}}(\ket{01}-\ket{10}).
  \end{align}
In the standard basis, the Bell's mixture can be written as:
  \begin{align}\label{eq-Bellsmixtures}
  \rho_B&=\lambda_{1}\ket{\Phi^+}\bra{\Phi^+}+\lambda_{2}\ket{\Phi^-}\bra{\Phi^-}\notag\\
        &+\lambda_{2}\ket{\Psi^+}\bra{\Psi^+}+\lambda_{4}\ket{\Psi^-}\bra{\Psi^-}
  \end{align}
where $\lambda_{i} \hspace{6pt}(i=1\sim4)$ are the eigenvalues of
the $\rho_B$.
  One may rewrite Eq.~\ref{eq-Bellsmixtures} in the spinor basis~\cite{SuTele,Su,SuTsai1,SuTsai2,SuTsai3,SuTsai4}:
  \begin{align}\label{eqwerner2qbt}
   \rho_B=\frac{1}{4}(I\otimes I+\sum^3_{i=1}\Omega_{ii}(-1)^{\epsilon_i}\sigma_{ii}),
  \end{align}
  here $\epsilon_{1}=\epsilon_{3}=0,\epsilon_{2}=1$, and
  \begin{align}\label{rel-eig2omega}
  &\lambda_{1}=\frac{1}{4}(1+\Omega_{11}+\Omega_{22}+\Omega_{33}),\notag\\
  &\lambda_{2}=\frac{1}{4}(1-\Omega_{11}-\Omega_{22}+\Omega_{33}),\notag\\  &\lambda_{3}=\frac{1}{4}(1+\Omega_{11}-\Omega_{22}-\Omega_{33}),\notag\\
  &\lambda_{4}=\frac{1}{4}(1-\Omega_{11}+\Omega_{22}-\Omega_{33}).
  \end{align}

  In the following we show the sufficient and necessary conditions for the separability of $\rho_B$ can be expressed by the inequality:
  \begin{align}\label{ineq-sep}
  1\leq\sum^3_{i=1}|\Omega_{ii}|
  \end{align}
  One can obtain the above inequality based on the method which is different from the PT~\cite{Horo2} or Wootters occurrence~\cite{Wootters}.

  First of all, we prove the inequality Eq.~\ref{ineq-sep} is a necessary condition for bipartite separability of $\rho_B$.
  Suppose $\rho_B=\sum^N_{k=1}p_k\rho^A_{k}\otimes\rho^B_{k}$ is separable,
  $p_k\geq 0$ and $\sum^N_{k=1}p_k=1$.
  In terms of spinor representation,
  the density operators $\rho^A_k$ and $\rho^B_k$
  are written as
 \begin{align}\label{eqDNAB2qbt}
   \rho^A_k=\frac{1}{2}\sum^3_{i=0}\Omega^A_{k,i}\sigma_i
   \hspace{2pt}\text{ and }\hspace{2pt}
   \rho^B_k=\frac{1}{2}\sum^3_{j=0}\Omega^B_{k,j}\sigma_j,
 \end{align}
  here $\Omega^A_{k,i},\Omega^B_{k,j}\in{\mathbb{R}}$ and $\Omega^A_{k,0}=\Omega^B_{k,0}=1$
  for $0\leq i,j\leq 3$.
  The state $\rho_B=\sum^N_{k=1}p_k\rho^A_{k}\otimes\rho^B_{k}$
  is thus rephrased as
 \begin{align}\label{eq-WstateRe-2qbt}
  \rho_B=\frac{1}{2}\sum^N_{k=1}\sum^3_{i,j=0}
   p_k\Omega^A_{k,i}\Omega^B_{k,j}\sigma_{ij}.
 \end{align}
  According to Eqs.~\ref{eqwerner2qbt} and~\ref{eqDNAB2qbt},
  we obtain the following relations, for $0\leq i,j\leq 3$,
  \begin{align}\label{eqrelas2qbt}
   &\sum^N_{k=1}p_k\Omega^A_{k,0}\Omega^B_{k,0}=1;\nonumber\\
   &\sum^N_{k=1}p_k\Omega^A_{k,i}\Omega^B_{k,i}=\Omega_{ii}\hspace{3pt}\text{ as }\quad i\neq 0;\nonumber\\
   &\sum^N_{k=1}p_k\Omega^A_{k,i}\Omega^B_{k,j}=\Omega_{ij}=0\hspace{2pt}\text{ as }\quad i\neq j.
  \end{align}
  These relations remain true for the general instance in the next section.
  Since both $\rho^A_k$ and $\rho^B_k$ are positive,
  the inequalities hold
  \begin{eqnarray}\label{eqtrace2qbt}
   {\rm Tr}^2(\rho^A_k)\geq{\rm Tr}((\rho^A_k)^2)
   \text{ and }
   {\rm Tr}^2(\rho^B_k)\geq{\rm Tr}((\rho^B_k)^2),
  \end{eqnarray}
  which leads to
  \begin{eqnarray}\label{eqOmega2qbt}
   (\Omega^A_{k,0})^2\geq\frac{1}{2}\sum^3_{i=0}(\Omega^A_{k,i})^2
   \text{ and }
   (\Omega^B_{k,0})^2\geq\frac{1}{2}\sum^3_{i=0}(\Omega^B_{k,i})^2.
  \end{eqnarray}
  By multiplying the inequalities of parties $A$ and $B$ of
  Eq.~\ref{eqOmega2qbt} and using the Cauchy's inequality,
  one acquires
  \begin{eqnarray}\label{eqCauchy2qbt}
   \Omega^A_{k,0}\cdot\Omega^B_{k,0}\geq\frac{1}{2}\sum^3_{i=0}|\Omega^A_{k,i}\cdot\Omega^B_{k,i}|.
  \end{eqnarray}
  Finally multiplying the weight $p_k$ to both sides of
  Eq.~\ref{eqCauchy2qbt} and summing over the $N$ terms,
  the following inequality is valid
  \begin{align}\label{eqans2qbt}
   &\sum^N_{k=1}p_k\Omega^A_{k,0}\Omega^B_{k,0}\notag\\
   &\geq
   \frac{1}{2}\sum^N_{k=1}\sum^3_{i=0}p_k|\Omega^A_{k,i}\Omega^B_{k,i}|
   \geq\frac{1}{2}\sum^3_{i=0}|\sum^N_{k=1}p_k\Omega^A_{k,i}\Omega^B_{k,i}|.
  \end{align}
  Through the relations of Eq.~\ref{eqrelas2qbt}, the necessary proof of the inequality Eq.~\ref{ineq-sep} are completed.

  In further, we prove the inequality Eq.~\ref{ineq-sep} is also a sufficient condition for bipartite separability of $\rho_B$. Suppose the inequality Eq.~\ref{ineq-sep}
  holds for the Bell's mixture $\rho_B$.
  We develop a separable form for the Bell's mixture
  \begin{align}\label{eqsepform2qbt}
   &\rho_B=\frac{1}{4}\sum^4_{l=1}\rho^A_l\otimes\rho^B_l\text{ with}\notag\\
   &\rho^A_1=\frac{1}{2}(\sigma_0+(-1)^{\epsilon_{1}}\sqrt{|\Omega_{11}|}\sigma_1 \notag\\
    &\hspace{33pt}+(-1)^{\epsilon_{2}}\sqrt{|\Omega_{22}|}\sigma_2+(-1)^{\epsilon_{3}}\sqrt{|\Omega_{33}|}\sigma_3),\notag\\
   &\rho^A_2=\frac{1}{2}(\sigma_0-(-1)^{\epsilon_{1}}\sqrt{|\Omega_{11}|}\sigma_1 \notag\\
    &\hspace{33pt}+(-1)^{\epsilon_{2}}\sqrt{|\Omega_{22}|}\sigma_2-(-1)^{\epsilon_{3}}\sqrt{|\Omega_{33}|}\sigma_3),\notag\\
   &\rho^A_3=\frac{1}{2}(\sigma_0+(-1)^{\epsilon_{1}}\sqrt{|\Omega_{11}|}\sigma_1 \notag\\
    &\hspace{33pt}-(-1)^{\epsilon_{2}}\sqrt{|\Omega_{22}|}\sigma_2-(-1)^{\epsilon_{3}}\sqrt{|\Omega_{33}|}\sigma_3),\notag\\
   &\rho^A_4=\frac{1}{2}(\sigma_0-(-1)^{\epsilon_{1}}\sqrt{|\Omega_{11}|}\sigma_1 \notag\\
    &\hspace{33pt}-(-1)^{\epsilon_{2}}\sqrt{|\Omega_{22}|}\sigma_2+(-1)^{\epsilon_{3}}\sqrt{|\Omega_{33}|}\sigma_3),\notag\\
   &\rho^B_1=\frac{1}{2}(\sigma_0+\sqrt{|\Omega_{11}|}\sigma_1-\sqrt{|\Omega_{22}|}\sigma_2+\sqrt{|\Omega_{33}|}\sigma_3),\notag\\
   &\rho^B_2=\frac{1}{2}(\sigma_0-\sqrt{|\Omega_{11}|}\sigma_1-\sqrt{|\Omega_{22}|}\sigma_2-\sqrt{|\Omega_{33}|}\sigma_3),\notag\\
   &\rho^B_3=\frac{1}{2}(\sigma_0+\sqrt{|\Omega_{11}|}\sigma_1+\sqrt{|\Omega_{22}|}\sigma_2-\sqrt{|\Omega_{33}|}\sigma_3),\text{ and}\notag\\
   &\rho^B_4=\frac{1}{2}(\sigma_0-\sqrt{|\Omega_{11}|}\sigma_1+\sqrt{|\Omega_{22}|}\sigma_2+\sqrt{|\Omega_{33}|}\sigma_3).
  \end{align}
  where,$ (-1)^{\epsilon_{i}}=sign(\Omega_{ii}),\hspace{2pt} i=1,2,3$. We show that each $\rho^A_l$ ($\rho^B_l$), $1\leq l\leq 4$ is a density operator
  if the inequality Eq.~\ref{ineq-sep} is satisfied.
  Obviously $\rho^A_l$ ($\rho^B_l$) are hermitian and
  have unit trace.
  It is easy to calculate the eigenvalues of each $\rho^A_l$
  ($\rho^B_l$) and
  there are only two kinds of eigenvalues
  \begin{eqnarray}\label{eqsegvalue2qbt}
   &\lambda^A_{l,1}=\lambda^B_{l,1}=\frac{1}{2}(\frac{2+\sqrt{4-4(1-|\Omega_{11}|-|\Omega_{22}|-|\Omega_{33}|)}}{2});\nonumber\\
   &\lambda^A_{l,2}=\lambda^B_{l,2}=\frac{1}{2}(\frac{2-\sqrt{4-4(1-|\Omega_{11}|-|\Omega_{22}|-|\Omega_{33}|)}}{2}).
  \end{eqnarray}
   These two eigenvalues are positive if $1\geq|\Omega_{11}|+|\Omega_{22}|+|\Omega_{33}|$
   and thus $\rho^A_l$ ($\rho^B_l$) are density operators. Therefore, if the inequality Eq.~\ref{ineq-sep} is satisfied, then $\rho_B$ is separable.

   Then, we operated Peres PT criterion:
  \begin{align}\label{eqwerner2qbt}
   \rho_B^{T_B}=\frac{1}{4}(I\otimes I+\sum^3_{i=1}\Omega_{ii}\sigma_i\otimes\sigma_i),
  \end{align}
  the eigenvalues of $ \rho_B^{T_B} $ are:
  \begin{align}\label{rel-eig2omega}
  &\lambda_{1}^{T_B}=\frac{1}{4}(1+\Omega_{11}-\Omega_{22}+\Omega_{33}),\notag\\
  &\lambda_{2}^{T_B}=\frac{1}{4}(1-\Omega_{11}+\Omega_{22}+\Omega_{33}),\notag\\  &\lambda_{3}^{T_B}=\frac{1}{4}(1+\Omega_{11}+\Omega_{22}-\Omega_{33}),\notag\\
  &\lambda_{4}^{T_B}=\frac{1}{4}(1-\Omega_{11}-\Omega_{22}-\Omega_{33}).
  \end{align}
  When $ \lambda_{i}^{T_B} \geq 0 \hspace{6pt} i=1\cdots4$ ,then $ \rho_B^{T_B} $ is separable. It should also be added that the conditions of $ \rho_B $ is a positive density operator $ \lambda_{i} \geq 0 \hspace{6pt} i=1\cdots4$. Therefor, one can obtain the same result as Eq.~\ref{ineq-sep} on the grounds that the inequalities both $ \lambda_{i} \geq 0$ and $ \lambda_{i}^{T_B} \geq 0$.

 \section{Bell Diagonal States of Higher Dimension}\label{secBellhighdim}
   In this section we show that the proof in the necessary condition
   can be extended to the more general occasion, a bipartite system
   ${\cal H}_A\otimes{\cal H}_B$ of dimension $2^p\times 2^p$.
   Extending the sufficient condition is difficult because it is
   not easy to find a separable form as of
   Eq.~\ref{eqsepform2qbt}.
   Thus we focus on the acquisition of the necessary condition
   for the bi-partite separability of Bell diagonal states.

 The Bell diagonal states in a $2^{p} \times 2^{p}$ system can be represented in the spinor basis as~\cite{SuTele,Su,SuTsai1,SuTsai2,SuTsai3,SuTsai4}:
 \begin{align}\label{eqWernerSp-2}
 \rho_{B}=\frac{1}{2^{2p}}\sum^3_{i_1,i_2,\cdots,i_p=0}(-1)^{\epsilon_{i_1}+\epsilon_{i_2}\cdots+\epsilon_{i_p}}\Omega_{i_1i_2\cdots i_p,i_1i_2\cdots i_p}
 \sigma_{i_1\cdots i_p,i_1\cdots i_p},
 \end{align}
 where $\epsilon_{{i_m}=0}=\epsilon_{{i_m}=1}=\epsilon_{{i_m}=3}=0$,
 $\epsilon_{{i_m}=2}=1$, $1\leq m\leq p$,
 and $\Omega_{i_1i_2\cdots i_p,i_1i_2\cdots i_p}=1$ if $i_1=i_2=\cdots=i_p=0$.
 We follow the same procedure as in the last section to acquire
 the necessary condition.
 Suppose the Bell diagonal states are bi-partite separable
 \begin{align}\label{eqSepWerner}
 &\rho_B=\sum^N_{k=1}p_k\rho^A_{k}\otimes\rho^B_{k}\text{ with}\notag\\
 &\rho^A_k=\frac{1}{2^p}\{\sum^3_{i_1,i_2,\cdots,i_p=0}\Omega^A_{k,i_1i_2\cdots i_p}\sigma_{i_1\cdots i_p}\}\text{ and}\notag\\
 &\rho^B_k=\frac{1}{2^p}\{\sum^3_{j_1,j_2,\cdots,j_p=0}\Omega^B_{k,j_1j_2\cdots j_p}\sigma_{j_1\cdots j_p}\},
 \end{align}
 here $p_k\geq 0$, $\sum^N_{k=1}p_k=1$, and $\Omega^A_{k,i_1i_2\cdots i_p},\Omega^B_{k,j_1j_2\cdots j_p}\in\mathbb{R}$.
 With Eqs.~\ref{eqWernerSp-2} and~\ref{eqSepWerner},
 we obtain the following relations as of Eq.~\ref{eqrelas2qbt}
 \begin{align}\label{eqrelateW}
  &\sum^N_{k=1}p_k\Omega^A_{k,00\cdots 0}\Omega^B_{k,00\cdots 0}=1,\notag\\
  &\sum^N_{k=1}p_k\Omega^A_{k,i_1i_2\cdots i_p}\Omega^B_{k,i_1i_2\cdots i_p}=\Omega_{i_1i_2\cdots i_p,i_1i_2\cdots i_p}\notag\\
  &\hspace{142pt} \text{ as } i_{1\leq r\leq p}\neq 0,\text{ and}\notag\\
  &\sum^N_{k=1}p_k\Omega^A_{k,i_1i_2\cdots i_p}\Omega^B_{k,j_1j_2\cdots j_p}=0\text{ for some } i_r\neq j_r.
 \end{align}
 By virtue of Eq.~\ref{eqtrace2qbt}, the coefficients obey the
 inequalities
 \begin{align}\label{eqOmega2pqbt}
   &(\Omega^A_{k,00\cdots 0})^2\geq\frac{1}{2^p}\sum^3_{i_1,i_2\cdots,i_p=0}(\Omega^A_{k,i_1i_2\cdots i_p})^2
   \text{ and}\notag\\
   &(\Omega^B_{k,00\cdots 0})^2\geq\frac{1}{2^p}\sum^3_{j_1,j_2\cdots,j_p=0}(\Omega^B_{k,j_1j_2\cdots j_p})^2.
  \end{align}
 By multiplying the inequalities of parties $A$ and $B$ of
  Eq.~\ref{eqOmega2pqbt} and using the Cauchy's inequality,
  one acquires
  \begin{align}\label{eqCauchy2pqbt}
   \Omega^A_{k,00\cdots 0}\cdot\Omega^B_{k,00\cdots 0}\geq\frac{1}{2^p}
   \sum^3_{i_1,i_2,\cdots,i_p=0}|\Omega^A_{k,i_1i_2\cdots i_p}\cdot\Omega^B_{k,i_1i_2\cdots i_p}|.
  \end{align}
 Similarly, multiplying the weight $p_k$ to both sides of
  Eq.~\ref{eqCauchy2pqbt} and summing over the $N$ terms,
 one derives the following inequalities
  \begin{align}\label{eqans2pqbt}
   &\sum^N_{k=1}p_k\Omega^A_{k,00\cdots 0}\Omega^B_{k,00\cdots 0}\notag\\
   &\geq
   \frac{1}{2^p}\sum^N_{k=1}\sum^3_{i_1,i_2,\cdots,i_p=0}p_k|\Omega^A_{k,i_1i_2\cdots i_p}\Omega^B_{k,i_1i_2\cdots i_p}|\notag\\
   &\geq\frac{1}{2^p}\sum^3_{i_1,i_2,\cdots,i_p=0}|\sum^N_{k=1}p_k\Omega^A_{k,i_1i_2\cdots i_p}\Omega^B_{k,i_1i_2\cdots i_p}|.
  \end{align}
  Complying with the relations of Eq.~\ref{eqrelateW},
  the inequality of Eq.~\ref{eqans2pqbt} leads to the required condition
  \begin{eqnarray}\label{eqans2pqbtx}
   1\geq \frac{1}{2^p}(\sum^3_{i_1,i_2,\cdots,i_p=0}|\Omega_{i_1i_2\cdots i_p,i_1i_2\cdots i_p}|),
  \end{eqnarray}
  which is a simple proof of necessary condition for bipartite separability of Bell diagonal states.

 \section{Conclusion}\label{secconclusion}
  In this article, we propose a new scheme to establish a criteria different from the PT and Wootters concurrence
  for the separability of Bell diagonal states.
  In a $2\times 2$ system,
  this scheme provides a sufficient and necessary conditions of
  the bi-partite separability for the Bell diagonal states and it could gives us an insight into quantum entanglement. When the separability of Bell diagonal states are transformed into entanglement,
  implied the local density operators $\rho^{\cal I}_k$ moved to the outside the Hilbert space ${\cal H}_{\cal I}$ for ${\cal I}=A,B$.
  This criteria, in general, in a $2^p\times 2^p$ system is simply a necessary condition for $p\geq 2$
  and a sufficient condition in the $2\times 2$ system.
  However, under the appropriate choices of separable forms, it is
  possible to modify that such criterion to obtained a sufficient condition of Bell diagonal states of arbitrary dimension.


\nonumsection{References}

 \end{document}